\PassOptionsToPackage{dvipsnames,table}{xcolor}

\documentclass[sigconf]{acmart}

\AtBeginDocument{%
  \providecommand\BibTeX{{%
    \normalfont B\kern-0.5em{\scshape i\kern-0.25em b}\kern-0.8em\TeX}}}

\copyrightyear{2025}
\acmYear{2025}
\acmConference[CHI '25]{CHI Conference on Human Factors in Computing Systems}{April 26-May 1, 2025}{Yokohama, Japan}
\acmBooktitle{CHI Conference on Human Factors in Computing Systems (CHI '25), April 26-May 1, 2025, Yokohama, Japan}\acmDOI{10.1145/3706598.3714112}
\acmISBN{979-8-4007-1394-1/25/04}

\definecolor{lightgray}{gray}{0.9}
\usepackage{balance}       
\usepackage{graphics}      
\usepackage[T1]{fontenc}   
\usepackage{color}
\usepackage{booktabs}
\usepackage{textcomp}
\usepackage{makecell}
\usepackage{xspace}
\usepackage{fancyvrb}
\usepackage{amsthm}
\usepackage{hyperref}
\usepackage{subcaption}
\usepackage{capt-of}
\usepackage{graphicx}
\usepackage{dcolumn} 
\usepackage{tipa}
\usepackage{varwidth}
\usepackage{tabularx}
\usepackage{caption}
\usepackage{color,soul}
\usepackage{geometry}

\usepackage{tikz}

\usepackage{microtype}        
\usepackage{ccicons}          

\newcommand{\eg}{\textit{e.g.}\@\xspace}
\newcommand{\ie}{\textit{i.e.}\@\xspace}
\newcommand{\etal}{\textit{et al. }}

\usepackage{multirow}
\usepackage{enumitem}
\setlist[description]{leftmargin=\parindent,labelindent=\parindent}

\begin{document}

\title{Beyond \textit{Omakase}: Designing Shared Control for Navigation Robots with Blind People}

\author{Rie Kamikubo}
\affiliation{
  \institution{University of Maryland, College Park}
  \state{Maryland}
  \country{United States}}
\affiliation{
  \institution{IBM Research - Tokyo}
  \city{Tokyo}
  \country{Japan}}
\email{rkamikub@umd.edu}

\author{Seita Kayukawa}
\affiliation{
  \institution{IBM Research - Tokyo}
  \city{Tokyo}
  \country{Japan}}
\email{seita.kayukawa@ibm.com}

\author{Yuka Kaniwa}
\affiliation{
  \institution{Waseda University}
  \city{Tokyo}
  \country{Japan}}
\email{ycrabring@ruri.waseda.jp}

\author{Allan Wang}
\affiliation{
  \institution{Miraikan - The National Museum of Emerging Science and Innovation}
  \city{Tokyo}
  \country{Japan}}
\email{allan.wang@jst.go.jp}

\author{Hernisa Kacorri}
\affiliation{
  \institution{University of Maryland, College Park}
  \state{Maryland}
  \country{United States}}
\email{hernisa@umd.edu}

\author{Hironobu Takagi}
\affiliation{
  \institution{IBM Research - Tokyo}
  \city{Tokyo}
  \country{Japan}}
\email{takagih@jp.ibm.com}

\author{Chieko Asakawa}
\affiliation{
  \institution{IBM Research}
  \state{New York}
  \country{United States}}
\affiliation{
   \institution{Miraikan - The National Museum of Emerging Science and Innovation}
   \city{Tokyo}
   \country{Japan}}
\email{chiekoa@us.ibm.com}

\renewcommand{\shortauthors}{Kamikubo et al.}

\begin{abstract}

Autonomous navigation robots can increase the independence of blind people but often limit user control---following what is called in Japanese an "omakase" approach where decisions are left to the robot. This research investigates ways to enhance user control in social robot navigation, based on two studies conducted with blind participants. The first study, involving structured interviews (N=14), identified crowded spaces as key areas with significant social challenges. The second study (N=13) explored navigation tasks with an autonomous robot in these environments and identified design strategies across different modes of autonomy. Participants preferred an active role, termed the "boss" mode, where they managed crowd interactions, while the "monitor" mode helped them assess the environment, negotiate movements, and interact with the robot. These findings highlight the importance of shared control and user involvement for blind users, offering valuable insights for designing future social navigation robots.

\end{abstract}



\maketitle

\section{Introduction}

With advancements in autonomous robot research and development, a growing number of robotic applications are being recognized, including those aimed for accessibility and social inclusion~\cite{hersh2015overcoming, han2024co, sahoo2024autonomous}. One actively explored use case is the development of navigation robots for blind people, with the hope to increase their independence and autonomy (\eg,~\cite{guerreiro2019cabot, tobita2017examination, azenkot2016enabling, cai2024navigating}). However, as technical advancements push the boundaries for fully autonomous robots, blind users may find themselves as followers~\cite{zhang2023follower}. This dynamic can be described as \textit{omakase} in Japanese~\cite{nihongo2024japanese}, an analogy also used in a decision-making model for leaving decisions to others (\eg, often to `experts' in the situation~\cite{akabayashi2001advanced,slingsby2004decision})---in this case, to the robot. In the context of disability, independence can be framed as control, choice, and power~\cite{rock1988independence}. Such \textit{omakase} robot interactions can create a critical design paradox for blind people, compromising independence when they have limited control over decisions about their environments~\cite{fine2005dependence,kane2009freedom}.

In this work, we aim to identify design strategies to enhance user control in robot navigation for blind people, with a focus on their choices of autonomy in complex real-world environments. While recent research has explored user control beyond the omakase (or follower) model---such as active control as a \textit{boss}~\cite{zhang2023follower} and companionship through \textit{monitoring} surroundings~\cite{hwang2024towards}---there is limited understanding of how these choices play out in social navigation scenarios. These environments, which involve nuanced social factors with people around, can constrain robot performance~\cite{mavrogiannis2023core} and make user control crucial for usability and overall acceptance~\cite{brewer2018understanding}. This work explores desirable human-robot interactions that support greater user control and agency for blind people in socially complex navigation scenarios.


To start our exploration, we conducted a preliminary study involving structured interviews with 14 blind participants to understand their challenges in social navigation and the strategies they currently use. Building on these insights, we designed a two-part user study with 13 blind participants (including 3 overlaps), first having them complete navigation tasks with an autonomous robot in real-world crowded scenarios informed by our initial interviews. 
These tasks were intended to prompt participants to engage with different modes of autonomy, which we introduced by adopting design metaphors that were culturally relevant: following the robot's movements passively as Omakase Mode, requesting spatial information as Monitor Mode, and making active commands as Boss Mode. We then conducted follow-up semistructured interviews to reflect on their experiences with these modes, gain a deeper understanding of their preferences, and brainstorm ideas for robot control and interaction. 

The results in this paper capture desirable user controls and interaction patterns with an autonomous robot. Findings from the navigation tasks revealed that blind participants often exercised Monitor Mode to assess the environment in crowded situations, followed by Boss Mode to play an active role in engaging with crowds based on information provided by the robot. The interaction between participants and the robot also involved negotiating movements, as participants attempted to understand the robot's capabilities and irregular movements. Follow-up interviews further revealed that participants were willing to take control of their interactions with bystanders by intervening in the robot's navigation. They expressed their desire to navigate more complex situations through mutual support and communication in user-robot interactions, such as the robot frequently explaining its movements, requesting human assistance, or issuing alerts on behalf of users when needed. The robot's role was discussed to be both supportive and collaborative.

The main contribution of this work is empirical. Through direct observation and reflection, we gained insights into user controls, commands, and interaction strategies from the perspective of blind participants using the autonomous navigation robot in real-world environments. Participants actively negotiated movements with the robot, switching between control modes like Monitor Mode to assess surroundings and Boss Mode to take charge of crowd interactions. This study highlights opportunities to develop interaction modes that allow users to retain agency in social navigation, particularly in complex environments, while still benefiting from the robot's capabilities. These insights can be used to shape future social robot designs that foster collaboration and trust between users and autonomous systems.
 
\section{Related Work}

There is a rich literature on autonomous navigation robots to support blind people, 
primarily equipped with capabilities for destination navigation~\cite{ranganeni2023exploring,zhang2023follower,Chen2023quadruped,Avila17},
obstacle avoidance~\cite{melvin2009rovi,ulrich2001guidecane,Hwang2023System}, a combination of both functions~\cite{tachi1984guide,wei2014new,wachaja2017navigating,guerreiro2019cabot,cai2024navigating}, and environmental recognition~\cite{wei2014new,Kuribayashi2023PathFinder}.
The underlying promise is that blind users can benefit from this assistance where they only need to follow the robot’s movements~\cite{guerreiro2019cabot} both outdoors~\cite{tachi1984guide,wei2014new,cai2024navigating,melvin2009rovi} and in indoor spaces~\cite{wachaja2017navigating,Kayukawa2023Enhancing,Kuribayashi2023PathFinder,Hwang2023System}. However, these movements are often unpredictable for blind users when following~\cite{guerreiro2019cabot} and may heighten feelings of \textit{alienation} if they cannot influence the interaction and outcomes~\cite{fronemann2022should}. In the context of broader navigation research, active user engagement has been discussed for its benefits~\cite{chrastil20121active, chrastil2014active,konishi2013spatial}.

In this section, we contextualize our work within the broader field of Human-Robot Interaction (HRI) as well as prior efforts in accessibility. Within the context of accessibility (Sec.~\ref{sec:RW_Human-Centered_Robotics}), we explore previous work that deviates from the typical robot-following approach and focuses on advancing control and autonomy. We situate these efforts within human-centered frameworks~\cite{grudin2018tool,shneiderman2022human} to discuss related design metaphors that can be applied in autonomous robot navigation for blind people. We then discuss literature on social robot navigation (Sec.~\ref{sec:RW_Social_Robot_Navigation}), bringing in unique challenges for the exploration of user control. In crowded environments with pedestrian traffic and queues, a robot’s behavior is determined by a balancing act between safety and social compliance. Such settings demand more opportunities for control and autonomy.

\subsection{Control in Human-Centered AI \& Robotics} 
\label{sec:RW_Human-Centered_Robotics}

The design of HRI and the integration of AI within these systems have been explored from various perspectives, with a growing emphasis on human-centered frameworks~\cite{sankaran2020respecting, shneiderman2022human}. Designers, developers, and other stakeholders are moving from the idea of autonomous agents to a more nuanced approach that considers human agency and collaboration~\cite{grudin2018tool,capel2023human}. Ben Shneiderman’s framework~\cite{shneiderman2022human} emphasizes metaphors like ``supertools'' and ``active appliances,'' positioning AI as a means to amplify, augment, and enhance human performance rather than replace it. This metaphorical approach is central to designing systems that maintain human control even in highly automated environments.
While full robot control can have its advantages for blind people, especially in unfamiliar environments~\cite{giudice2008blind}, the lack of control on technology and its assistance may create dependency on others when it fails~\cite{kane2009freedom}. This is particularly concerning given that blind people already have limited access to spatial information, which poses significant barriers to independence~\cite{banovic2013uncovering,giudice2018navigating}. To address these challenges, studies have suggested an interdependence framework that supports collaborative relationships between blind people and their surroundings~\cite{jain2023want} or assistive technologies~\cite{bennett2018interdependence}.


Our research builds on the existing knowledge of human control and design metaphors in human-centered frameworks, applying these concepts to the context of blind users and autonomous navigation robots. Efforts exploring this dimension with blind people (\eg, ~\cite{zhang2023follower,ranganeni2023exploring}) have shown that balancing automation with human control is crucial for fostering users’ confidence, comfort, and trust. Zhang \etal~\cite{zhang2023follower} also introduced the unique perspective of \textit{``I am the follower, also the boss''} reflecting blind users' experiences with different modes of autonomy with robotic companions. A similar dynamic can be seen in the roles of guide dogs and blind people~\cite{hwang2024towards}. While guide dogs can be considered \textit{``the eyes that lead''}~\cite{tucker1984eyes}, both parties engage in a collaborative effort for orientation and mobility~\cite{wiener2010foundations}. For example, blind people are responsible for general orientation---monitoring their surroundings and making navigational decisions by issuing cues (\ie, commands) to their guide dogs~\cite{hwang2024towards}. Similarly, blind users interacting with autonomous robots require a degree of control, such as the ability to intervene in decision making or adjust the robot’s behavior. 

No single mode of autonomy is optimal in every situation~\cite{ranganeni2023exploring}, making this balance between automation and user control critical.  Despite the ongoing efforts to improve control for blind people interacting with autonomous robots~\cite{zhang2023follower,ranganeni2023exploring}, identifying interaction mechanisms is still underexplored, especially in the context of social robot navigation. It involves complex settings that demand attention to social norms and the user's sense of agency.

\subsection{Social Robot Navigation}
\label{sec:RW_Social_Robot_Navigation}

There are many challenges in building and deploying robot navigation technologies in human environments. The survey paper by Mavrogiannis \etal~\cite{mavrogiannis2023core} categorizes the challenges into planning, behavioral, and evaluation challenges. Among the three, planning and behavioral challenges are closer to this work.

From a planning perspective, researchers see these challenges with the need to balance efficiency and safety~\cite{mavrogiannis2023core, francis2023principles}. When safety is prioritized, planners often encounter the ``freezing robot problem''~\cite{trautman2010freeze}, where robots stop unnecessarily even when navigable paths exist. This occurs because safety prioritization leads to inflated obstacle zones around pedestrians, and all viable paths in the robot’s planning space are blocked. To address this issue, researchers eacho that human-robot interaction modeling is essential~\cite{trautman2010freeze} and have leveraged learning-based methods~\cite{chen2017cadrl, liu2021srnn}, classic model-based approaches~\cite{ferrer2013socialforce, singamaneni2021cohan}, or hybrid techniques~\cite{wang2022mpcgroup, xiao2023mpctransformer}. Despite the recent advances, the ``freezing robot problem'' is an ongoing challenge~\cite{mavrogiannis2023core}. 

Our work is situated within the social navigation challenges, including the ``freezing robot problem,'' by exploring the concept of control and investigating how blind users of guide robots can engage with this issue. We also delve into the social behavior aspect of human-robot interaction. Existing research in this domain has studied a wide variety of behaviors, such as pedestrian group dynamics \cite{wang2020groups}, proxemic zones~\cite{hall1963proxemics, bachiller2022proxemics}, robot escape strategies~\cite{drazen2015childrenescape}, expression of robot intention~\cite{lichtenthaler2012intention}, and line waiting~\cite{nakauchi2000line}. However, this remains an area lacking established principled knowledge to guide robot design in social navigation~\cite{mavrogiannis2023core}. 

To address this gap, we broaden our understanding of social behaviors in navigation through a different lens. Numerous studies have examined blind people's experiences in navigating real-world environmentse~\cite{brunet2018strategies,williams2013pray,williams2014just, branham2017someone}, providing valuable insights to inform the design of technology that can support and enhance these experiences. Understanding their navigation strategies has also guided the development of assistive technologies~\cite{cattaneo2011blind,jeamwatthanachai2019indoor}. In this work, we adopt a similar methodology; as an initial step, we aim to understand blind people's social navigation experiences and strategies, with the goal of applying these insights to the design of social navigation robots with greater user control.

\section{Preliminary Interviews}
Leveraging a human-centered design approach, we draw on human-human interactions as key references to inform the design of human-robot interactions~\cite{shneiderman2022human}. We conducted a preliminary study with blind people using structured interviews to understand their social navigation experiences and strategies in everyday situations. Insights were intended to guide the design of our user study involving navigation tasks with an autonomous robot, exploring their choices of control in complex real-world environments.
Given this purpose, we framed this preliminary study around two key questions: (i) What challenges do you encounter when navigating around people, such as in crowded environments? and (ii) What strategies do you use to mitigate these challenges? Responses to the first question informed the social navigation scenarios for the tasks, while the second provided insights into interaction strategies for designing social robot navigation (introduced as probes for user commands in Sec.~\ref{sec:prompts}).

\subsection{Recruitment and Participants}

Our studies were approved by our institution's Institutional Review Board (IRB).
To advertise our preliminary study to the local blind community, we used (a) an existing mailing list associated with the museum in Japan---our collaboration institution included in the IRB and (b) visitor trials of the autonomous navigation robot deployed at the museum, where participants were invited for a separate interview. 
We designed the study to be structured with two focused questions and brief demographic questions, aiming to last no more than 20 minutes, and called for volunteer participation. We obtained their consent prior to audio recording their interview responses.

We recruited a total of 14 blind participants (6 women, 8 men), aged between 32 and 75 years (M = 52.79, SD = 12.15), with 12 participating remotely and 2 in person. Nine participants were totally blind, and five were legally blind. 
Three participants were blind since birth, and four were blind from a young age with the onset varying between ages 3 and 10. All participants were white cane users, and there was no participation from blind people who use guide dogs. This might be due to the small community in Japan, which is reported to have about 1/10 of the number of guide dogs available in the U.S.~\cite{nippon2020japan}. 

\subsection{Findings}
We used a deductive thematic analysis approach to examine participants' responses~\cite{braun2006using}. This method involved reviewing and organizing the responses based on predefined themes that focused on challenges and strategies related to social navigation, as derived from the focused interview questions. Through an analysis for recurring patterns and similarities, we identified two overarching scenarios that reflect major navigation challenges and workarounds to address these challenges in each context.

\textit{(1) Streams of People.} Participants reported difficulties in navigating crowded spaces, which they described as `streams of people.' This complexity was reported in various scenarios, including train stations, riverfront trails, hospitals, busy shopping districts, festivals, and other events. A key issue was their struggle to adapt to the unpredictable behaviors of the crowd, making it hard to tell whether others were walking toward or away from them, stopping, or forming a line. Being swept along by the crowd and feeling physically close to strangers was particularly frightening, often resulting in a loss of direction and orientation. Participants also mentioned that relying on audio cues, such as those from station gates or escalators, becomes less effective in crowded areas. 

Maintaining a safe distance was challenging as well, as participants could inadvertently trip others with their white canes or risk damaging the cane. In dense crowds, the limited ability to move the cane from side to side complicates obstacle detection and maintaining a straight path. Although we initially anticipated that this might be different for those who use guide dogs---typically trained to maintain a straight line and avoid obstacles~\cite{tucker1984eyes}, they too can experience challenges when navigating through streams of people. A participant from our main study later reported that crowds often fail to notice guide dogs, leading to safety issues due to the unpredictable behaviors of people around.

Participants shared various workarounds to navigate these situations. One common strategy was \textbf{asking others} for assistance. They also used localization techniques, such as \textbf{following physical cues} like tactile paving or \textbf{stopping temporarily} to gather information. \textbf{Following environmental cues}--like sounds from nearby shops or escalators, or even smells and wind--were considered helpful for navigation. Additional strategies included walking along walls or listening for reflective sounds, such as changes in echoes or audio announcements, to avoid collisions and identify crowded spaces. While this awareness of the crowds was intended for caution, participants also reported \textbf{following crowds} (\eg, people heading toward an escalator after getting off a train) as a way to find their direction. However, it was often described as a frantic process, where blind people were not intentionally recognizing the crowd but simply moving along with others as they were in close proximity. \textbf{Alerting others}, such as by tapping a white cane, was also a technique used to interface with the crowd, though it was not ideal.

\textit{(2) Lines.} Participants expanded on difficulties with identifying and following people in line. They reported uncertainty about where to stand, how the line was forming, and whether it was moving, especially in more complex queues like U-shaped or forked lines. They also found it difficult to maintain social distance and follow social rules. There was concern about unintentionally cutting in line or getting in the way of others who were already in line. These challenges were noted in various contexts, including waiting for buses, taxis, trains, escalators, ATMs, cash registers, restrooms, hospitals, and food distribution shelters.

Workarounds for standing in line and tracking its movement included \textbf{asking others} in line, similar to how they navigate through streams of people. They also asked store staff for information about the end of the line or to notify them when it was their turn. To gauge the line’s movement and positioning, participants reported \textbf{following audio cues} from footsteps or conversations of people in line. Assistive technology also played a role, including sonar devices and remote sighted services. To help move forward in the line, a white cane was used to gauge the distance between people in front. However, it often led to accidental contact and was considered a last resort when other methods were ineffective. They ultimately ask for sighted assistance, particularly in settings like hospitals or grocery stores, where standing in line is expected and there is a greater need for smooth queuing and navigation.

\section{User Study: Navigation Tasks and Semistructured Interviews}

This next study is an exploration including navigation tasks and semistructured interviews to elicit blind participants' preferences and needs for control with a navigation robot. Building on the social navigation challenges identified in our preliminary interviews, the navigation tasks focus on two key scenarios: streams of people and lines in real-world environments. These scenarios are particularly challenging for both humans and robots due to the `freezing robot problem'~\cite{trautman2010freeze} and potential misalignment with social behaviors.  During the tasks, participants explore strategies for navigating these situations, using different modes of autonomy called Omakase, Monitor, and Boss (defined in Sec.~\ref{sec:Procedure}). We also probe participants on strategies they could employ, referring to them as commands such as 'alerting' or 'following,' based on insights from the preliminary interviews. After completing the tasks, participants reflect more broadly on what actions they would take in such scenarios and what support they would expect from the robot to enhance their social navigation strategies and experiences.

To design our study and enable participants to experience the social navigation scenarios, we conducted our study at a museum where an autonomous robot has been deployed to assist visitors, primarily blind people. We chose the museum environment for its real-world complexity and relevance, with an anticipated average of 2969 visitors per day during the study period, based on statistics from August 2023.


\subsection{Recruitment and Participants}
\label{sec:user_study_participants}

\begin{table*}[t]
    \centering
    \small
    \caption{Demographic information of user study participants.}
    \label{tab:participants}
    \Description{
    This table shows the demographic of participants. There are 13 participants listed in this table. P01 (Man) is a 73-year-old participant who has been legally blind with only light perception for over 10 years. He uses a cane as a mobility aid and travels independently a few times a week. P02 (Woman) is a 49-year-old participant who has been totally blind for over 10 years. She uses a cane as a mobility aid and travels independently 5 times a week. P03 (Man) is a 54-year-old participant who has been totally blind for over 10 years. He uses a cane as a mobility aid and travels independently every day. P04 (Woman) is a participant who has had low vision and no peripheral vision for over 10 years. She primarily uses a guide dog and occasionally a cane as mobility aids, and travels independently 5 times a week. P05 (Woman) is a 51-year-old participant who has been legally blind with only hand motion for 5 years. She uses a cane as a mobility aid and travels independently 4 times a week. P06 (Woman) is a 60-year-old participant who has been totally blind for over 10 years. She uses a cane as a mobility aid and travels independently a few times a month. P07 (Woman) is a participant who has been totally blind for over 10 years. She uses a cane as a mobility aid and travels independently 5 times a week. P08 (Man) is a 58-year-old participant who has been legally blind with only light perception for over 10  years. He uses a cane as a mobility aid and travels independently 2-3 times a week. P09 (Woman) is a 44-year-old participant who has been totally blind in her left eye and had no peripheral and color vision in her right eye for over 10  years. She uses a cane as a mobility aid and travels independently every day. P10 (Man) is a 58-year-old participant who has been totally blind for over 10  years. He uses a cane as a mobility aid and travels independently every day. P11 (Woman) is a 59-year-old participant who has been totally blind in her right eye and had some light perception in her left eye for over 10  years. She primarily uses a guide dog and occasionally a cane as mobility aids, and travels independently every day. P12 (Man) is a 42-year-old participant who has had low vision and no peripheral vision for 3  years. He uses a cane as a mobility aid and travels independently 5 times a week. P13 (Woman) is a 60-year-old participant who has been legally blind with no peripheral vision for over 10  years. He uses a cane as a mobility aid and travels independently every day.
    }
    \begin{tabular}{lllllll}
\toprule
 \multirow{2}{*}{ID} & \multirow{2}{*}{Age} & \multirow{2}{*}{Gender} & \multirow{2}{*}{Vision Level} & Impairment& Mobility Aid(s) & Solo Travel\\ 
 & & & & Duration& (*: primary) & Frequency\\ 
        \midrule
P1 & 73  & man   & legally blind; light perception        & >10 years & white cane       & few times/week  \\ 
P2 & 49  & woman & totally blind                          & >10 years & white cane       & 5 times/week    \\ 
P3 & 54  & man   & totally blind                          & >10 years & white cane       & everyday        \\ 
P4 & n/a & woman & low vision; no peripheral vision       & >10 years & guide dog* & 5 times/week    \\
   &     &       &                                        &           & \& white cane    &                 \\
P5 & 51  & woman & legally blind; hand motion             & 5 years   & white cane       & 4 times/week    \\ 
P6 & 60  & woman & totally blind                          & >10 years & white cane       & few times/month \\ 
P7 & n/a & woman & totally blind                          & >10 years & white cane       & 5 times/week    \\ 
P8 & 58  & man   & legally blind; light perception        & >10 years & white cane       & 2--3 times/week \\ 
P9 & 44  & woman & totally blind (left);                  & >10 years & white cane       & everyday        \\ 
   &     &       & no peripheral and color vision (right) &           &            &                 \\
P10 & 58 & man   & totally blind                          & >10 years & white cane       & everyday        \\ 
P11 & 59 & woman & totally blind (right);                 & >10 years & guide dog* & everyday        \\ 
    &    &       & some light perception (left)           &           & \& cane    &                 \\
P12 & 42 & man   & low vision; no peripheral vision       & 3 years   & white cane       & 5 times/week    \\ 
P13 & 60 & woman & legally blind; no peripheral vision    & >10 years & white cane       & everyday        \\ 
        \bottomrule
\end{tabular}
\end{table*}

We recruited participants through a mailing list of nearly 200 blind or low vision people. 
The eligibility criteria were as follows: people who regularly navigate with a white cane or guide dog; aged 18 and older; ability to go to the study site (the entrance of the museum) by themselves or with others \eg, sighted guides. They were compensated 7500 yen for their time (2.5 hours) and transportation expenses.
After screening the initial responses to the call for participation from 19 subscribers, we invited a total of 13 participants, as listed in Table~\ref{tab:participants}.
Among these participants, three experienced the navigation robot for the first time. Among those with experience, the number of times they experienced the robot ranged from one to six.
While nine participants have visited the study location in the past (either as personal visits to the museum or other research study visits), no one reported being familiar with the environment to navigate on their own using their mobility aid.
P3, 10, and P13 also participated in the preliminary interviews.
As participants with other impairments, P8 had dwarfism and slight difficulty in bending their joints; and P12 had mild hearing loss and used a hearing aid.

\subsection{Apparatus}

\begin{figure*}[h]
    \centering
\includegraphics[width=\textwidth]{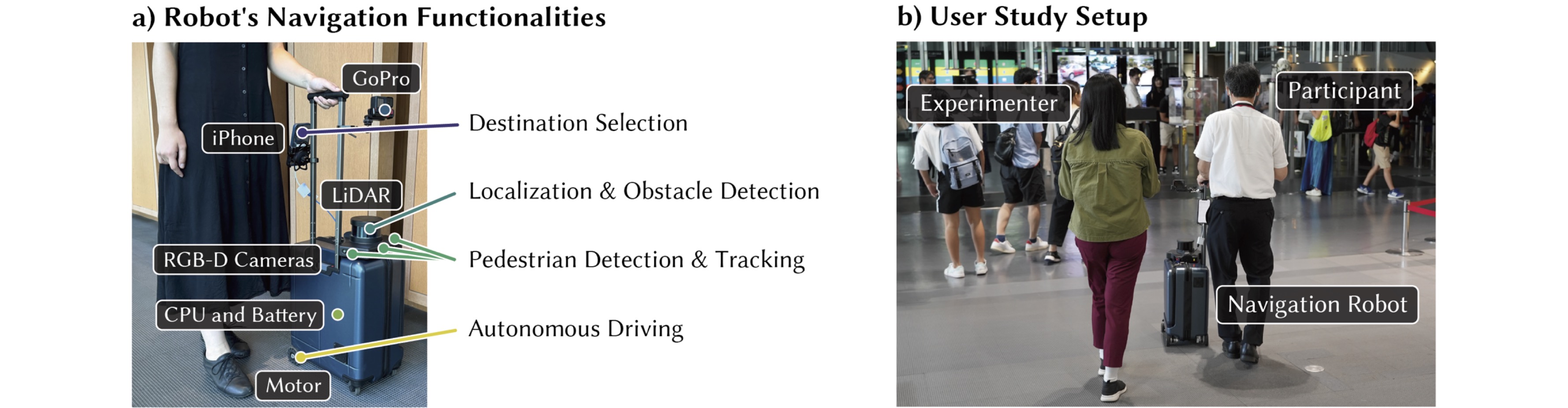}
    \caption{a) The robot's key functionalities in navigation and b) the user study setup with the robot, participant, and experimenter}
    \label{fig:hardware_and_setup}
    \Description{
    This figure displays two real images. The left image shows the robot used in our study, which is suitcase-shaped. An iPhone and a GoPro are mounted on the handle, used for controlling the robot and recording the experiment, respectively. A LiDAR sensor is mounted on top of the suitcase for localization and obstacle detection, along with three RGB-D cameras for human detection. The suitcase contains computational resources such as a CPU and battery, with motorized rear wheels. The right image depicts a blind participant walking through a museum with the robot. An experimenter follows behind to interact with the participant.
    }
\end{figure*}

We used an autonomous navigation robot deployed in the real world for our exploration. The robot’s hardware and software are based on an open-source project~\cite{cabot-github}.
Computational resources, including a CPU and battery, are housed in a commercially available suitcase, with motorized rear wheels (Figure~\ref{fig:hardware_and_setup}a). 
The robot localizes its position and orientation by matching point clouds from its LiDAR sensor to a pre-built map. 
Once the robot’s destination is set via a smartphone, the navigation planner calculates the least-cost path to reach it.
During navigation, the robot detects surrounding obstacles using the LiDAR sensor and plans a safe path to avoid them. 
Additionally, the robot tracks nearby people with three RGB-Depth cameras powered by the YOLOv4 image recognition engine~\cite{bochkovskiy2020yolov4}, ensuring it maintains an appropriate distance from individuals. 
Upon reaching the destination, an announcement is made through the smartphone. 
To record the environment during the experiment, we mounted a GoPro camera on the suitcase handle.


\subsection{Study Design}
\label{sec:Procedure}
\begin{figure*}[t]
    \centering
    \includegraphics[width=\textwidth]{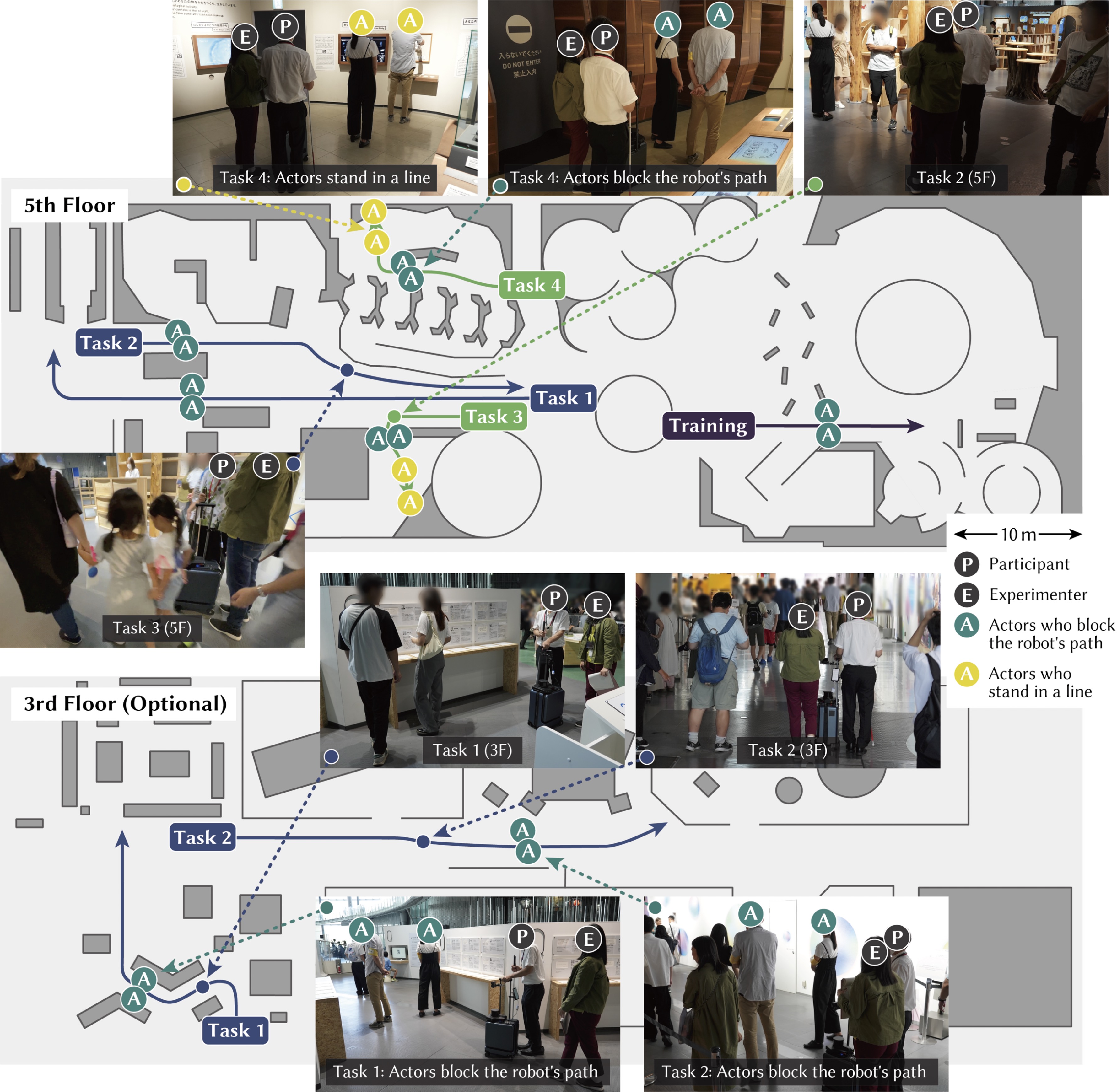}
    \caption{
    Routes used in the navigation tasks, where participants navigated a crowded environment in the museum with the robot and experimenter. Actors either blocked the robot’s path or stood in a line at the destination.}
    \label{fig:main_study_routes}
    \Description{
    This figure illustrates the routes walked by participants during the experiment on the floor maps of the museum (3rd and 5th floors). On the 5th-floor map, five routes (Training, Task1, Task2, Task3, Task4) are depicted: Training: A straight route of about 20 meters, during which two actors block the robot’s path; Task1 and Task2: Straight routes of about 40 meters through the exhibition area, where two actors block the path; Task3 and Task4: Routes of approximately 20 meters, where participants enter a specific exhibition area and move toward a designated exhibit. Two actors block the path, and two more actors stand in line at the goal in front of the exhibit.
    On the 3rd-floor map, two routes, Task1 and Task2, are shown. Along these routes, two actors also block the robot’s path.
    The figure includes four photos depicting the participants, the experimenter, and two actors: 1) Task 1 on the 3rd floor, where actors block the robot’s path. 2) Task 2 on the 3rd floor, where actors block the robot’s path. 3) Task 4 on the 5th floor, showing actors blocking the robot’s path. 4) Task 4, where the participant approaches actors standing in line.
    Additionally, four photos show interactions between the participants, the robot, and museum visitors (who are not actors) during the experiment: 1) A family of four crosses in front of the participant. 2) Two visitors walk across the participant’s path. 3) Two people stand in front of the participant, viewing an exhibit. 4) A crowded scene, with many people walking around the participant.
    }
\end{figure*}

\subsubsection{Prompts and Probes.}
\label{sec:prompts}
Before starting the navigation tasks, we prompted participants with three modes of autonomy and asked them to use a think-aloud method to communicate their choices of modes during the tasks. This approach was designed to reflect their desired levels of engagement and decision-making. We built on related design metaphors from prior literature, specifically drawing from more culturally aligned perspectives from Zhang \etal~\cite{zhang2023follower}'s follower and boss roles for blind people interacting with robots. These roles can be further understood through the social robot design dimension~\cite{axelsson2021social}, which includes a leadership model that determines ``Who initiates the interaction? Who determines what happens next?'' This model captures a spectrum from robot-led to user-led, with a mutual role reflecting dynamics similar to the relationship between guide dogs and blind people. In this mutual effort, blind people monitor their surroundings to make navigational decisions~\cite{hwang2024towards}. The following explains three distinct modes of autonomy, based on these roles, for interacting with the environment and the robot: Omakase Mode, Monitor Mode, and Boss Mode.

\begin{itemize}
    \item \textbf{Omakase} - Characterized by a low level of autonomy, this mode involves a passive role with minimal decision-making by the participant. For example, this mode is selected when the participant follows the robot’s decisions without actively engaging in the navigation process, even if the robot stops due to the freezing problem.
    \item \textbf{Monitor} - Characterized by a developing level of autonomy, this mode involves the participant engaging with their environment by asking questions or seeking information from the robot. For example, this mode is selected when the participant asks questions such as ``Are there people blocking the way?'' or ``How many people are there?'' to guide their navigation choices. In this study, this mode is realized using the Wizard-of-Oz method, where the experimenter responds to user queries based on a predefined set of information the robot can detect, such as pedestrians and obstacles~\cite{cabot-github}. We adopted this method to encourage participants to reflect on the interaction aspects of this level of autonomy, rather than focusing on the technical aspect. AI-driven conversational models have been explored to assist blind people with scene understanding, calling for human-like Q\&A systems~\cite{gamage2023blind,bigham2010vizwiz}. However, users often find responses unreliable due to missing or incorrect information, leading to reduced trust~\cite{Kaniwa2024ChitChatGuide,clark2019makes}. Introducing such systems is logistically challenging, and addressing these technical limitations is beyond the focus of our study. 

    \item \textbf{Boss} - Characterized by a high level of autonomy, this mode involves the participant taking full control of their navigation and employing strategies. This mode is annotated when participants issue commands such as ``Alert people'' or ``Move forward'' and interact directly with bystanders by saying ``Excuse me'' to navigate through crowded areas. The participant’s role in directing or choosing these actions is central to this mode.
\end{itemize}

Additionally, before the navigation tasks, we probed examples of commands in Boss Mode based on our preliminary interviews that revealed blind people's strategies in social navigation. Participants were encouraged to brainstorm their own commands, with the provided list serving only as a guide. 

\begin{itemize}
    \item \textbf{User Inquiry/Alert} - Participants ask or alert specific people to move aside or to make their presence known to those around them, such as saying ``Excuse me. I'm coming through.''
    \item \textbf{Robot Inquiry/Alert} - Participants instruct the robot (which in this study was the experimenter) to ask or alert specific people to move aside, regardless of the method, such as beeping or saying ``Excuse me.''
    \item \textbf{Movement} - Participants direct the robot's movements, such as waiting and moving forward, which often conflict with the robot's predefined movement rules.
    \item \textbf{People Following} - Participants instruct the robot to follow specific people.
    \item \textbf{Physical Cue Following} - Participants instruct the robot to follow physical features such as a wall or tactile paving.
\end{itemize}

\subsubsection{Procedure.} We first explained the purpose of the study, obtained informed consent, and conducted a training session prior to the user tasks.
During the training, participants first walked with the robot without taking any actions, experiencing Omakase Mode, and adjusted the robot's walking speed.
Then, to explain what participants can do in the other two modes, we created a situation where the robot couldn’t move because its path was blocked by two actors.
In this situation, they were instructed to ask or negotiate with the robot to understand their surroundings (Monitor Mode) and to make a command to clear the path (Boss Mode).

The robot used in this study can navigate fully autonomously to its destination while avoiding pedestrians and obstacles along the way. It can also wait in line autonomously near the destination. However, in crowded scenarios, the safety-priority design often causes the robot to stop, which often triggers users to switch modes as discussed in Sec.~\ref{sec:prompts}.

\textbf{Navigation Tasks.}
Participants were asked to walk along four specified routes during regular business hours, freely switching between three modes until they arrived at each destination. 
Figure~\ref{fig:main_study_routes} shows the robot's regular paths (task 1 through task 4), which participants followed in numerical order.
Of the four routes, two (task 1 and task 2) were routes to experience stream situations and the other two (task 3 and task 4) were routes to experience queuing situations.
Although the user tasks were primarily conducted on the fifth floor of the museum, task 1 and task 2 were performed on the third floor when the fifth floor was not crowded.
On each route, two actors completely blocked the robot's path at one point, and for the routes of task 3 and task 4, they also created a line in front of the destination.
To artificially create situations where the robot can move in a line, we selected exhibits as the destinations for task 3 and task 4, where the robot is prohibited from calculating a detour due to the narrowness in front. 
During the tasks, the experimenter walked one step behind the robot and interacted with participants and surroundings on behalf of the robot in Monitor and Boss Modes (Figure~\ref{fig:hardware_and_setup}b).
The experimenter responded to participants' questions in Monitor Mode based on predetermined rules for explaining the surroundings, such as ``People (or obstacles) are blocking the path (or have just crossed the path).''
After completing task 2 or task 4, participants took brief interviews to facilitate post-session interviews.
They were asked about any difficult situations they faced during tasks, what they could do in such situations, and what kind of support they would like from the robot.

\textbf{Post-session Interviews.}
After completing all of their tasks, participants took post-session semistructured interviews.
First, they were asked open-ended questions about how they would resolve challenging situations, including both those they mentioned in brief interviews during the tasks and situations they hadn’t experienced, such as meandering lines or single-file queues, which were prepared in advance based on preliminary interviews.
They freely bounced around ideas about what they wanted to do, such as possible commands they could issue and what kind of support they needed, including specific information, in each situation.
Second, they were also asked to answer a set of questions on a seven-point scale with reasons, which is reported in Figure~\ref{fig:likert_overall} and Figure~\ref{fig:likert_user_or_robot-led}.
Q1--Q3 were Likert-scale questionnaires about each mode (1: strongly disagree, 4: neutral, 7: strongly agree), and Q4--Q8 asked how they preferred the leadership for each command (1: fully robot-led, 4: neutral, 7: fully user-led).
Lastly, they were asked about their demographic information, detailed in Sec.~\ref{sec:user_study_participants}.
  
\subsection{Analysis}

Our data analysis was structured around two main phases of the study: user navigation tasks and post-session interviews. We transcribed audio recordings from both phases and used audio and video recordings from the navigation tasks to capture user behavior, interactions, and crowd dynamics. 

\subsubsection{User Tasks Annotations}
\label{sec:analysis_user_tasks_annotations}
We employed a multi-faceted approach to analyze user tasks, annotating the modes of autonomy and issued commands while mapping them to our prompts and probes listed in Sec.~\ref{sec:prompts}. We also annotated user queries in Monitor Mode and crowd situations during the interaction events, as described below. The coding scheme was refined iteratively within the team. One researcher took the lead on annotations, while at least one other researcher conducted a detailed pass. Any annotation uncertainties were resolved in team meetings or offline discussions.

\textit{Monitor queries.}
We delved into Monitor Mode where users query the robot for information. Our goal is to find out what types of information the users seek in Monitor Mode. We employed a descriptive and qualitative analysis coding procedure to develop annotations. The procedure is a manual iterative clustering-based procedure over two steps. In the first step, we read through the participants' queries and generated concepts (\textit{a.k.a.} information request types). In the second step, we assigned the concepts back to each query. If a query could not be clustered into existing concepts appropriately, we created a new concept and reexamined all previous queries with the updated set of concepts. We repeated the two-step process until all queries could be mutually exclusively clustered into these concepts. Finally, we made sure that a consensus was reached on the final set of concepts and their assignments to the queries. These concepts are our annotations.
\begin{itemize}
    \item \textbf{Obstacles} - Queries about something that participants sensed was blocking the robot's path.
    \item \textbf{People} - In-depth queries about the status of the pedestrians around the robot.
    \item \textbf{Action} - Queries related to the robot's current action.
    \item \textbf{Attempt} - Queries on whether the robot can try something.
    \item \textbf{Situation} - General queries about what is happening or if there are any problems.
    \item \textbf{Line} - Queries related to line formations.
    \item \textbf{Destination} - Queries related to the navigation destination.
\end{itemize}

\textit{Crowd situation descriptions.} 
We grounded our coding of crowd situations on established social navigation literature. 
In two recent survey papers~\cite{mavrogiannis2023core, francis2023principles}, the most prevalent categorization to divide social navigation scenarios is whether the crowds are stationary or dynamic. 
It is commonly accepted that static crowds behave distinctively differently from dynamic crowds, and separate rules need to be considered for social navigation algorithms. 
Additionally, we noted the need to separate line-related scenarios from static crowd situations as a special case. 
This is also based on established literature~\cite{nakauchi2000line, kuribayashi2021linechaser}, which ascertains that crowd behavior in lines is unique and is difficult to tackle by a generic social navigation algorithm. 
In summary, we coded crowd situation descriptions according to the following:
\begin{itemize}
    \item \textbf{Stationary} - All pedestrians in the robot's close proximity were stationary (\eg, holding conversations, watching exhibits, looking at phones). 
    \item \textbf{Dynamic} - All pedestrians in the robot's close proximity were walking in any direction. 
    This also implies that soon after the subjects had initiated the interaction, the pedestrians were likely no longer around.
    \item \textbf{Line} - The subject and the robot arrived at the start of the line or were moving in the line.
    \item \textbf{Other} - Occasionally, interactions were initiated when no humans were around. 
    This typically happened when the robot failed to plan its course promptly.
\end{itemize}
Note that there were rare scenarios where both static and dynamic humans were present. 
In this case, the responses gathered were used in the analysis of both static and dynamic cases, because we believe the participants' behavior was influenced partially by both types of pedestrians.

\subsubsection{Post-Session Interviews}

The interviews included both open-ended questions for design discussions and Likert scale questions with accompanying shorter qualitative responses. 

For the qualitative analysis of the open-ended discussions, we employed inductive thematic analysis~\cite{braun2006using}. The process was led by researcher R1 who facilitated all discussion sessions. R1 began by assigning descriptive (\ie, semantic) codes to the transcripts of participant comments and feedback. These codes were then organized into categories. R1 proceeded to generate initial themes and sub-themes based on these codes, integrating interpretations throughout. In parallel, researcher R2 conducted the analysis of Likert scale responses. R2 looked into the spread and variability of responses using medians to represent central tendencies and interquartile ranges (IQR). R2 also summarized how participants selected each Likert scale option and identified general trends in their comments.

The preliminary findings from both analyses were refined through team meetings for a more unified understanding of the concepts in the data~\cite{braun2021one}. R1 and R2 complemented each other's analyses; for example, R1 identified an initial theme related to \textit{user involvement} in human-robot interactions, which was broadened with R2's summary of responses concerning independence and control. In the following section, we present our findings from the interviews, focusing on participants' perspectives on autonomy and shared control. These insights are contextualized by the user tasks they performed with the autonomous robot prior to the interviews, which helped them reflect on their interactions and discuss future interfaces.
\section{Results}

\subsection{Choices for Autonomy and User Queries \& Commands in Navigation Tasks}
\label{sec:results_overall}

In this section, we report our observations of how participants explored and transitioned between different interaction modes for autonomy to complete navigation tasks in various crowd situations (Sec. \ref{sec:results_mode_transition}), as well as their queries (Sec.~\ref{sec:result_Monitor_Conversation}) and commands (Sec.~\ref{sec:result_Boss_Command}) to exercise more control outside Omakase Mode.

\subsubsection{Crowd Situation and Mode Transition Analysis} 
\label{sec:results_mode_transition}
\begin{figure*}
    \centering
    \includegraphics[width=\textwidth]{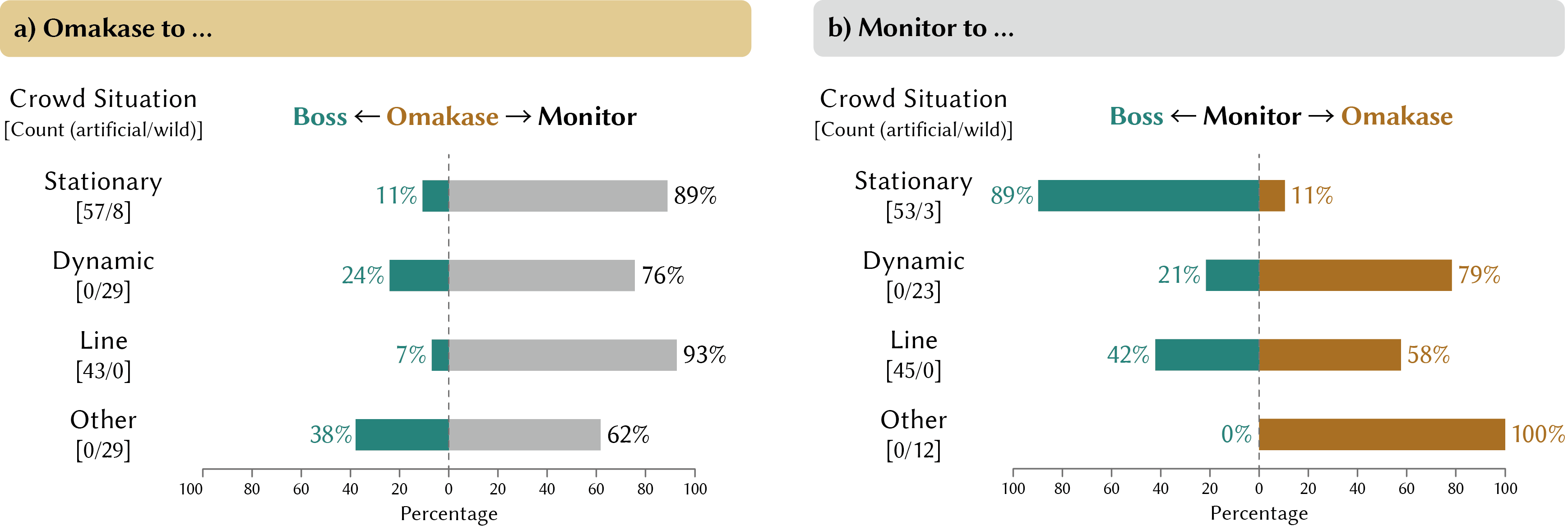}
    \caption{
    Percentage of transitions between modes of autonomy by crowd situations: a) from Omakase Mode to Boss or Monitor Mode and b) from Monitor Mode to Boss or Omakase Mode.}
    \label{fig:state_transition_rate}
    \Description{
    This figure is divided into two sections.
    Left section shows a graph of the percentage of transitions from Omakase mode to either Boss or Monitor mode for each Crowd Situation.
    In the Stationary situation, there were 57 artificial instances caused by actors and 8 wild instances involving real visitors. 11\% transitioned to Boss mode, while 89\% transitioned to Monitor mode.
    In the Dynamic situation, there were 29 wild instances, with 24\% transitioning to Boss mode and 76\% to Monitor mode.
    In the Line situation, there were 43 artificial instances, with 7\% transitioning to Boss mode and 93\% to Monitor mode.
    In the Other situation, there were 29 wild instances, with 38\% transitioning to Boss mode and 62\% to Monitor mode.
    Right section shows the percentage of transitions from Monitor mode to either Boss or Omakase mode for each Crowd Situation.
    In the Stationary situation, there were 53 artificial instances and 3 wild instances. 89\% transitioned to Boss mode, while 11\% transitioned to Omakase mode.
    In the Dynamic situation, there were 23 wild instances, with 21\% transitioning to Boss mode and 79\% to Omakase mode.
    In the Line situation, there were 45 artificial instances, with 42\% transitioning to Boss mode and 58\% to Omakase mode.
    In the Other situation, there were 12 wild instances, with 100\% transitioning to Omakase mode.
    }
\end{figure*}

Figure~\ref{fig:state_transition_rate}a shows the ratio of transitions to Monitor or Boss Mode from Omakase Mode.
We observed that participants often transitioned to Monitor Mode from Omakase Mode, such as asking for information about their surroundings or the reason the robot stopped. These situations involved the robot’s path being blocked (Stationary: 89\%), a few passersby walking across the robot’s path (Dynamic: 76\%), and people forming lines (Line: 93\%). Other situations (62\%) also led participants to switch to Monitor Mode, partly due to feedback and interface disruptions that made it difficult to understand why the robot stopped, such as audio feedback delays for indicating destination arrival or the deactivation of the touch sensor on the handle when participants were not consciously gripping it.  

When participants transitioned to Boss Mode directly from Omakase Mode, we observed two main types of situations. First, participants, having already understood the environment as crowded or detected people nearby based on surrounding sounds, issued commands directly to address the robot's movements, such as it zigzagged unpredictably or stopped for a long time.
This was particularly common in stationary or line situations.
Second, some participants (notably P10) constantly gave verbal alerts (\eg, "Excuse me") to their surroundings while walking, regardless of the situation. This led to an increased frequency of transitioning into Boss Mode in dynamic and other situations.

Figure~\ref{fig:state_transition_rate}b shows the ratio of transitions to Omakase or Boss Mode from Monitor Mode, indicating participants' choice of autonomy after gaining spatial information.
Participants tended to shift to Boss Mode in stationary situations (89\%) and to Omakase Mode in dynamic and other situations (79\% and 100\%, respectively).
In stationary situations, the trend of choosing Boss Mode was not surprising, as the robot's path was blocked (often by the actors in our study), leaving participants with no option but to ask them to step aside.
In dynamic and other situations, Omakase Mode was optimal, as the robot’s path was eventually cleared with no further action needed. For example, P7 asked, \textit{``What's wrong?''} to which the experimenter responded, \textit{``The robot tried to take a detour because people had just crossed the path.''} P7 shifted to Omakase Mode after assessing the situation.

In line situations, the ratio of transitions to Omakase Mode from Monitor Mode was slightly higher than that to Boss Mode. 
This was because after assessing the line situation upon reaching the end of the line, participants chose to follow the robot's movement without issuing any commands. Yet, there were instances where participants gave commands to the robot to follow the person in line.
Additionally, when the line moved, some participants queried the current status of the line, such as the number of people waiting, but typically returned to Omakase Mode without giving further commands.
For example, when P12 arrived at the end of the line, they asked, \textit{``Is this a line, or are people just standing in front?''} and the experimenter responded, \textit{``There are people standing in front of us, these are people waiting in line.''} P12 followed up with \textit{``How many people are in line?''} to which the experimenter responded, ``Now there are two people in front of us in line.'' P12 then confirmed, \textit{``OK, I will shift to Omakase Mode.''}

\subsubsection{User Query Analysis} 
\label{sec:result_Monitor_Conversation}

\begin{figure*}[t]
    \centering
    \includegraphics[width=\textwidth]{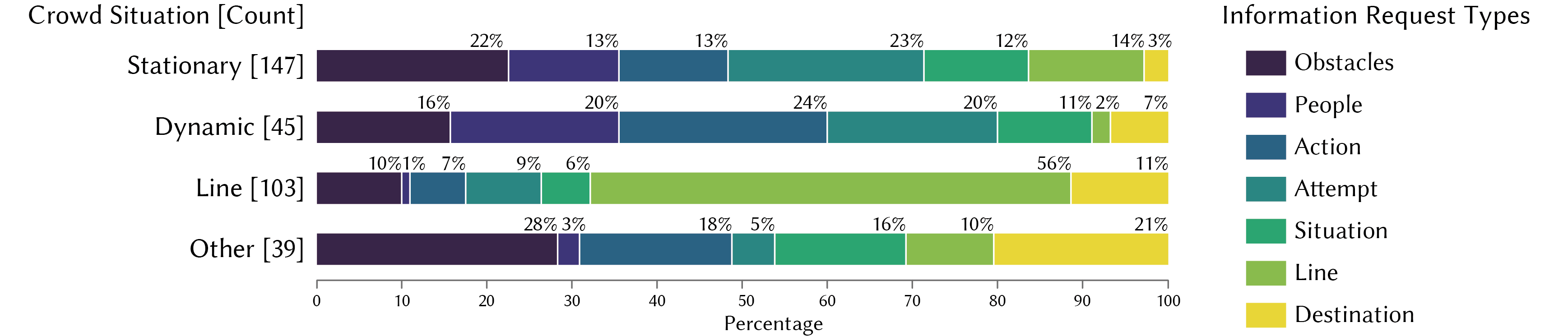}
    \caption{
    Percentage of information requests by crowd situations in Monitor Mode.}
    \label{fig:user_request_types_rate}
    \Description{
    This figure shows a graph representing the proportion of Information Request Types selected by participants in Monitor mode for each crowd situation. 
    In the Stationary situation, there were 147 samples, with Obstacle being selected 22\% of the time, People 13\%, Action 13\%, Attempt 23\%, Situation 12\%, Line 14\%, and Destination 3\%. 
    In the Dynamic situation, out of 45 samples, Obstacle was chosen 16\% of the time, People 20\%, Action 24\%, Attempt 20\%, Situation 11\%, Line 2\%, and Destination 7\%. 
    For the Line situation, there were 103 samples, with Obstacle selected 10\% of the time, People 1\%, Action 7\%, Attempt 9\%, Situation 6\%, Line 56\%, and Destination 11\%. 
    In the Other situation, with 39 samples, Obstacle was chosen 28\% of the time, People 3\%, Action 18\%, Attempt 5\%, Situation 16\%, Line 10\%, and Destination 21\%.
    }
\end{figure*}

\begin{figure*}[h]
    \centering
    \includegraphics[width=\textwidth]{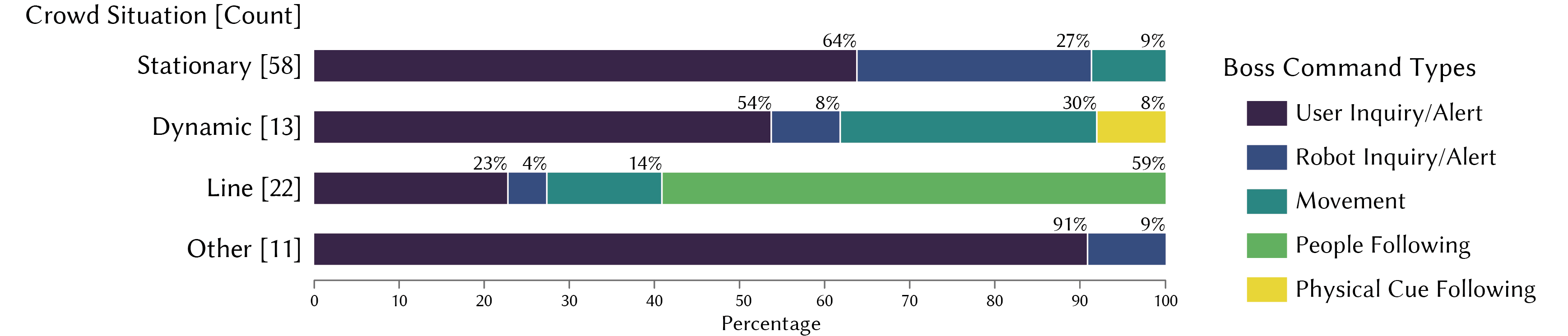}
    \caption{
    Percentage of user commands by crowd situations in Boss Mode.}
    \label{fig:boss_command_percentage}
    \Description{
    This figure shows a graph representing the proportion of Boss commands selected by participants for each crowd situation. 
    In the Stationary situation, there were 58 samples, with User Inquiry/Alert being selected 64\% of the time, Robot Inquiry/Alert 27\%, and Movement 9\%.
    In the Dynamic situation, there were 13 samples, with User Inquiry/Alert being selected 54\% of the time, Robot Inquiry/Alert 8\%, Movement 30\%, and Physical Cue Following 8\%. 
    In the Line situation, there were 22 samples, with User Inquiry/Alert being selected 23\% of the time, Robot Inquiry/Alert 4\%, Movement 14\%, and People Following 59\%. 
    In the Other situation, there were 11 samples, with User Inquiry/Alert being selected 91\% of the time, and Robot Inquiry/Alert 9\%. 
    }
\end{figure*}

We analyzed what types of queries participants made under different crowd situations in Monitor Mode. We observed varying distributions of information request types conditioned on each of the four crowd situations during Monitor Mode, as shown in Figure~\ref{fig:user_request_types_rate}. 

In stationary crowd scenarios, participants mostly queried about obstacles (22\%) and tried to explore what the robot could attempt in such situations (23\%). When making obstacle-type queries, participants often sought confirmation, such as \textit{``Is there something in front?''} In other cases, participants inferred that the robot had stopped due to obstacles and asked for more details, such as \textit{``What is in front?''}
Regarding attempt-type queries, when stopped, participants asked if the robot could try taking a detour to the left or right, or if it could proceed forward through the blocking obstacles or people. This might indicate that participants used Monitor Mode to explore the robot’s capabilities and negotiate strategies for navigating the obstacles. 
Another inquiry type relevant to the stationary crowd situation was about the line (14\%). In waiting-in-line tasks, participants were unsure if the robot had stopped because of the line, prompting questions such as \textit{``Is this a line?''} or \textit{``Are people waiting in line or just standing around?''}

In dynamic situations, participants made distinctively different types of queries. Action-type queries were dominant in these situations (24\%) compared to stationary situations (13\%). This happened often due to the robot taking zigzagging paths when navigating in dynamic crowds, which prompted the participants to question its actions, such as \textit{``Why are you turning right?''} or `\textit{`Why can't we go straight?''} Interestingly, we see a higher percentage of people-type queries (20\%) compared to those made in stationary situations (13\%). They could sense when pedestrians were no longer present nearby and asked to confirm with questions such as \textit{``Did someone just pass by?''} On a related note, in dynamic situations, obstacle-type queries were less common (16\%), as participants often assumed the obstruction was caused by people in the surroundings. Therefore, they mostly inquired for more details about people, asking \textit{``Is it crowded?''} or \textit{``How many people are there?''}

Unsurprisingly, line-type queries dominated the line crowd situations (56\%). Upon closer inspection, most participants wanted to confirm whether they were in the line (35\% of line-type queries). Other common queries included \textit{``How many people are in the line?''} (17\%), \textit{``Are we at the start of the line?''} (12\%), \textit{``Are we at the end of the line?''} (10\%), and \textit{``Let me know when the line moves''} (9\%). At times, participants didn't query and directly issued commands to the robot to follow the line.

``Other'' crowd situations typically took place due to occasional technical failures or the robot taking time to chart its course, which often confused participants. This confusion is reflected in the increased amount of obstacle-type queries (\eg, \textit{``What is in front?''} 28\%), action-type queries (\eg, \textit{``Why did you stop?''} 18\%), situation-type queries (\eg, \textit{``What is the situation?''} 16\%), and destination-type queries (\textit{``Have we arrived?''} 21\%).

\subsubsection{User Commands Analysis} 
\label{sec:result_Boss_Command}

We analyzed the commands participants issued in different crowd situations during Boss Mode. Figure~\ref{fig:boss_command_percentage} shows the ratio of each command type across these situations.

Inquiry/alert commands were often made in stationary and dynamic situations (91\% and 62\%, respectively), with user inquiry/alert commands being particularly preferred in both situations (64\% and 54\%, respectively). This meant participants actively engaged with passersby to clear the way, such as by using direct alerts \textit{``I'm coming through''} or \textit{``Please move to the side''}, or by asking more politely \textit{``Could you move out of the way please?''}
Robot inquiry/alert commands involved speaking similar phrases on behalf of users and making a beep sound.
In dynamic situations, movement commands were also common (30\%). These commands typically involved asking the robot to move in a specific direction, wait until the crowd cleared, or return to a 
point where it had been moving smoothly and reroute.

In line situations, people-following commands were the most frequently used (59\%). 
Robot-led movement commands related to standing in line were also made (14\%).
User inquiry/alert and robot inquiry/alert commands happened when participants mistakenly instructed people to move aside, not realizing they were actually in line. 

Other notable commands included physical cue-following commands, which we observed by P11 issuing the instruction to \textit{``walk along the edge of the walkway''} to feel relieved by touching the wall. P10 frequently made verbal alerts regardless of surroundings, resulting in the ratio of user inquiry/alert commands accounting for 91\% in ``other'' situations.

\subsection{Perspectives on Autonomy and Interfaces for Control} 

\begin{figure*}
    \centering
    \includegraphics[width=\textwidth]{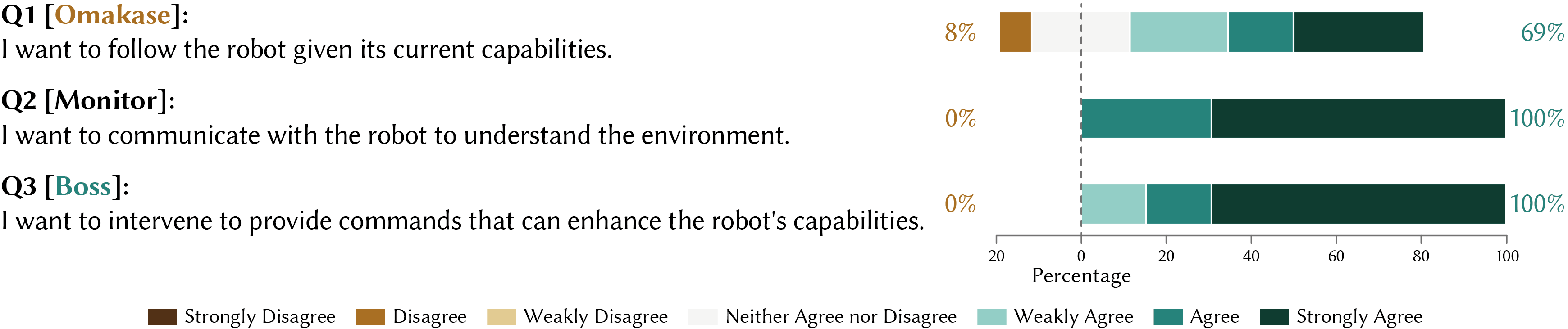}
    \caption{
    Overall experience ratings toward different modes of autonomy.
    }
    \label{fig:likert_overall}
    \Description{
    This figure illustrates the questions and results of three items answered using a 7-point Likert scale. Participants rated their responses on a scale where 7 points indicate “strongly agree,” 6 points indicate “agree,” 5 points indicate “weakly agree,” 4 points indicate “neutral,” 3 points indicate “weakly disagree,” 2 points indicate “disagree,” and 1 point indicates “strongly disagree.”
    Q1 is a question about Omakase mode with the statement, “I want to follow the robot given its   current capabilities.” 8\% of participants gave a score of 2, while 69\% of respondents gave a score of 5 or higher.
    Q2 is a question about Monitor mode with the statement, “I want to communicate with the robot to understand the environment.” All of participants gave a score of 5 or higher.
    Q3 is a question about Boss mode with the statement, “I want to intervene to provide commands that can enhance the robot’s capabilities.”
    All of participants also gave a score of 5 or higher.
    }
\end{figure*}

\begin{figure*}
    \centering
    \includegraphics[width=\textwidth]{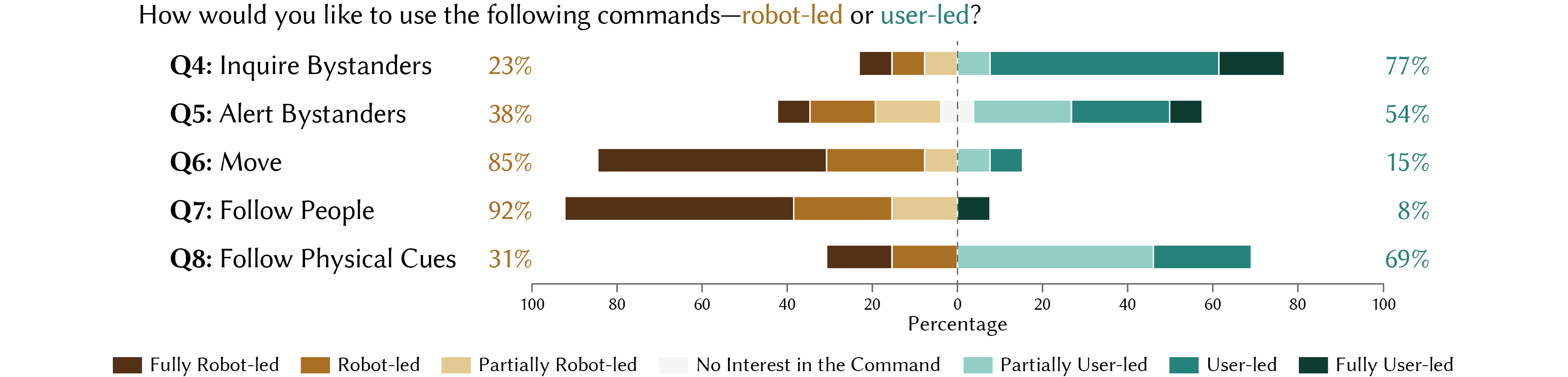}
    \caption{
    Participants' preference for commands on a scale from fully robot-led to fully user-led, with a midpoint indicating no interest in the command.
    }
    \label{fig:likert_user_or_robot-led}
    \Description{
    This figure illustrates the distribution of responses to five questions (Q4-Q8) asking users whether they preferred robot-led or user-led control for each Boss command. Responses were given on a 7-point Likert scale, where 1 represented “fully robot-led,” 2 “robot-led,” 3 “partially robot-led,” 4 “neutral,” 5 “partially user-led,” 6 “user-led,” and 7 “fully user-led.”
    For Q4, the command was “Inquire Bystanders,” with 23\% of participants selecting 3 or below, and 77\% selecting 5 or above. For Q5, the command was “Alert Bystanders,” with 38\% choosing 3 or below, and 54\% choosing 5 or above. For Q6, the command was “Move,” with 85\% selecting 3 or below, and 15\% selecting 5 or above. In Q7, the command was “Follow people,” with 92\% selecting 3 or below, and 8\% selecting 5 or above. Finally, in Q8, the command was “Follow physical cues,” with 31\% selecting 3 or below, and 69\% selecting 5 or above.
    }
\end{figure*}
In this section, we reflect on participants' discussions on autonomy and interaction design for greater control in autonomous robot navigation.
We also report together the trends in their autonomy choices and interaction preferences based on the questionnaire results. Figure~\ref{fig:likert_overall} and Figure~\ref{fig:likert_user_or_robot-led} summarize the results from the Likert scale questions on participants' preferences for each mode (Figure~\ref{fig:likert_overall}, Q1--Q3) and preferences for using boss commands in either a robot-led or user-led manner (Figure~\ref{fig:likert_user_or_robot-led}, Q4--Q8).
Although 9 out of 13 participants agreed that they want to follow the robot (Omakase Mode) given its current capabilities (Q1), all participants expressed a preference for using Monitor Mode to understand the surrounding environment (Q2) and Boss Mode to enhance the robot's capabilities (Q3). This section explores in detail how participants envision interacting with these modes. 

\subsubsection{User vs Robot Involvement Beyond Omakase}
\label{sec:result_user_robot_involvement}
One recurring theme in the interviews was the active role blind participants sought to play in social navigation, rather than being entirely dependent on the robot. A key factor steering this perception was their sense of agency---having control over their actions and outcomes~\cite{moore2016sense}. For example, P13 said, \textit{``As long as we understand the situation, humans should make the decisions. I think it's better for humans to make the decisions, rather than leaving everything to robots, because after all, people should be the ones in control.''}
As a result, all participants agreed that they want to intervene to provide commands that can enhance the robot's capabilities (Q3 in Figure~\ref{fig:likert_overall}).
Four participants (P4, P11--P13) mentioned that the reason was because they wanted to take an active role in the decision-making process, rather than leaving everything to the robot.
Regarding the command-related questions (Q4--Q8, Figure~\ref{fig:likert_user_or_robot-led}), participants who preferred user-led control referred to user autonomy as the reason for their preference.

Participants recognized that some complex social navigation situations require more user involvement, and direct communication with bystanders is crucial. This includes scenarios like standing in a multi-line queue or interacting with a crowd with specific needs. For example, P12 mentioned navigating shared spaces with individuals using strollers, canes, or wheelchairs, where direct communication is necessary to negotiate a clear path.: \textit{``It can be hard to ask them to move aside...It’s about having a bit of communication where both parties adjust slightly to create a path for each other.''} 
Opting for direct communication also reflected skepticism about whether the robot could fully understand social complexities or communicate effectively on behalf of users. P4 said: \textit{``Because even if you let the robot make the decision, I wonder if it would really understand... For example, like the line we mentioned earlier, if you ask the robot, `Is this the line for such and such?' and it says, `No,' you move on, and if it says, `Yes,' you still can’t be sure if it’s really a line or not. In the end, I think it's better to just ask the people around you directly.''}
Therefore, human intervention was considered a more efficient and reliable strategy in social navigation. A similar trend was observed in responses to questions regarding commands for inquiring/alerting bystanders (Q4 and Q5 in Figure~\ref{fig:likert_user_or_robot-led}). 
For Q4 (inquire bystanders), the median response was 6, with 10 out of 13 participants preferring user-led control (rating 5 or higher).
For Q5 (alert bystanders), the median response was 5, with seven participants favoring user-led control.
Five participants (P2, P7--P10) explained that their preference was due to the user's ability to handle complex situations more smoothly than the robot.

Additionally, participants' concerns about social acceptance contributed to their preference for human interaction. It was more appropriate for them to engage with bystanders, as opposed to relying on the robot to interface with the crowd such as through alerts. One participant (P10) explained this sentiment: \textit{``I think the robot shouldn’t issue warnings unless there’s a sense of urgency. I don't feel I have the right to assert myself that much.''}
This highlights the perception that using the robot for alerts or crowd interactions might come across as assertive in everyday social contexts. This is also reflected in participant responses to Q5 (alert bystanders), where the majority preferred user-led control, with four participants (P2, P3, P6, P11) expressing concerns about social acceptance.

Nevertheless, participants described robot involvement in navigation tasks as effective in certain situational contexts. For example, P13 appreciated the robot's navigation technology: \textit{``Sometimes, I just don’t feel like asking for help. That’s why having a robot that can move forward on its own would be amazing.''} P3 also highlighted situations where user communication might fail: \textit{``If people don't hear me or are engaged in other activities and don't move aside, and considering my limited ability to speak foreign languages, I thought it might be helpful if the robot could use warning sounds or announcements like `I am coming through' or `Please move aside.'}
Others like P4, P5, P10, and P11 echoed this sentiment, emphasizing that in crowded environments, robots generating non-intrusive alert sounds can effectively draw attention and facilitate smoother passage, as P5 stated, \textit{``it would be less embarrassing if the robot played music or said something like `passing through' instead of me calling out myself.''} 
Participants like P1 and P12 further supported this idea, suggesting different alert options depending on the area's crowd density. P12 explained: \textit{``If there are just one or two people, a simple alert to get them to look and move aside would be enough. However, in busier places where many paths overlap, even if an alert goes off, people might not realize it’s directed at them. They could just ignore it, thinking it's a random sound like a smartphone going off,''} proposing that the robot should say `AI suitcase is passing through' in areas like train stations. 
While non-visual feedback, such as beep sounds, is actively explored for guiding blind people~\cite{kayukawa2019bbeep,rodriguez2012assisting}, we found that participants were more interested in a mixed approach with both user and robot involvement. As P6 suggested: \textit{``Before I even have to say anything as the user, the robot could automatically make a sound like `beep beep' to issue a warning. That way, I could also become aware that there’s someone there through the sound, and then I could say, `Excuse me, could you let me through?'''}

\subsubsection{Boss Interaction and Robot Support}
\label{sec:comment_boss_interaction_robot_support}
Participant feedback on the boss interactions in social robot navigation exposed several strategies, including those demonstrated in the navigation tasks under Boss Mode (Sec.~\ref{sec:result_Boss_Command}). Participants envisioned active user involvement by asking or alerting bystanders themselves (\eg, saying \textit{`Excuse me'} or inquiring \textit{`Where is the end of the line?'}). We also identified their need for support from the robot to complement their interactions with other people, such as locating them in their surroundings (\eg, ``In which direction are they?''). For example, P11 shared their experience: \textit{``Figuring out whether it's a person or not is the most crucial thing. Quite often, we end up saying things like `sorry' to a pole, and we apologize, even to walls, saying `I'm sorry, excuse me,' no matter what we bump into.''}

Additionally, we collected various comments related to the types of information needed about people nearby---often to ask for help if necessary. These included proximity (P4: \textit{``If they're within 50 centimeters, I might ask them directly''}), activity level (P13: \textit{``If they're walking, they might just pass by but if they're standing still, there’s a possibility they could respond.''}), orientation \& attention (P7: \textit{``I can’t tell if they’re facing away or looking toward me, so I’d like to know that.''}), and age \& ability to communicate (P8: \textit{``Are they in a certain age group? Also, whether or not they can speak <the same language>, perhaps''}). 
In their navigation tasks under Monitor Mode (Sec.~\ref{sec:result_Monitor_Conversation}), we similarly observed people-type queries related to status (\textit{``Have they moved or not?''}), proximity (\textit{``How far are they?''}), activities (\textit{``What are they doing?''}), and attention (\textit{``Have they noticed us?''}). 

While such information about nearby pedestrians was often intended to facilitate interactions with the crowd, it was also used to avoid direct engagement. 
Aligning with blind people's information need relating to children~\cite{kamikubo2024we}, P11 explained, \textit{``Children are at a lower height and their movements can be unpredictable. Much like when I'm with guide dogs, the mindset changes depending on whether children are present or not...this awareness can be helpful.''} P1 also made a query about children in Monitor Mode during the navigation tasks (\eg, \textit{``Are there children?''}).

Strategies for movement, also observed in the navigation tasks under Boss Mode, were discussed for flexible navigation through crowds, reflecting a unique relationship between the blind user and the robot. This relationship is collaborative: the robot provides essential feedback, while the user maintains control. For example, P8 highlighted that the robot could monitor and report the situation, while the user would decide when to proceed or halt: \textit{``If the obstacle is moving, you could activate the `wait mode' if the robot is moving left and right, unable to find a clear path. And if the obstacle is stationary, you could `monitor' it and switch to `boss'. I think you'll probably be able to choose between those options.''} In another example, P6 explained: \textit{``I think it’s possible to ask which route is more open (to the robot) and use that information to give instructions for a detour.''}  
Such communication for movement strategies resembles the dynamics of collaboration between guide dog handlers and their companions~\cite{hwang2024towards}.

Following cues that could be detected by the robot (\eg, movements of other people) or the user (\eg, tactile paving) was another important strategy. This was not only for navigating complex social environments but also for enhancing safety and comfort, as reflected in the use of the ``people following'' and ``physical cue following'' commands in Boss Mode (Sec.~\ref{sec:result_Boss_Command}). 
They envisioned instructing the robot to follow the last person in line or maintain an appropriate social distance while waiting in line, hoping to overcome the difficulties as described by P13: \textit{``What I always find troublesome is not being able to see the movements of the person in front of me. I move forward, saying `Excuse me' with my cane...but then I end up bumping into things and apologizing. Sometimes it's such a hassle that I ask the person in front if I can hold onto their arm to help me get through, but doing this every time is tiring.''}
The ``physical cue following'' command, such as following tactile paving or wall edges, was also deemed critical for comfort, as it would help participants walk straight in crowded spaces and alert others to move aside with ease.

\subsubsection{Monitor Interaction and Mutual Support}
\label{sec:comment_monitor_interaction_mutual_support}
A recurring theme centered around dialogue and negotiation in participants' discussions about interacting with robots. 
This often involved mutual support to address complex social environments, similar to their interactions with sighted guides providing situational updates.
For instance, when moving through narrow spaces, P1 described how they communicate with the guide to adjust positions: \textit{``With a sign like this, you can move through narrow spaces. It would be nice if a robot could do something similar, like saying `Please walk behind.'''} 
In another example, P11 reflected on the robot making a strange movement and indicated their need for communication, shifting away from Omakase Mode:\textit{``When you can't see and the robot makes some strange movements, it can make you feel uneasy, like, `What's happening?'...If the robot said, `There are a lot of people, so I’ll move in a zigzag pattern,' or just gave a simple explanation like that, you’d understand, `Oh, the robot is moving like this to avoid people, that’s why it’s zigzagging.' Then, there would be no anxiety about the movement.''}

Given these comments, it is not surprising that all participants preferred Monitor Mode to understand their surroundings (Q2 in Figure~\ref{fig:likert_overall}). 
P7 rated 7 in Q2 provided the following notable comment: \textit{``By being aware of the situation, people can understand things like, 'Oh, that's why the robot was wandering around earlier,' or 'It stopped because someone was standing in the way.' I think everyone wants to know what those who can see are seeing. If we can grasp that, it feels like walking won't just be walking anymore—it'll become something more enjoyable.''}

We also found that mutual support between blind users and robots involved users taking the initiative to assist the robot in navigating complex situations. For example, P13 described how the robot could prompt users to seek help from others when it reaches its limited capabilities, suggesting, \textit{``The robot could tell me, `I don’t know what to do from here, so ask the person behind you,' or `The line splits from here, and I’m not sure where to line up, so ask the person behind you.''}' 
This would encourage the user to step in, such as asking other people \textit{``Excuse me, does the line split from here?''}. Similarly, P5 highlighted the importance of the robot communicating difficulties in crowded environments, proposing that if the robot said, \textit{``It's a little difficult to pass through right now''} the user might think, \textit{``Oh, I need to help.''} 

Additionally, participants' views on interaction with the robot revealed a nuanced understanding of broader societal connections, including concerns that excessive reliance on robots might lead to diminished human engagement. P4 said, \textit{``Even if robots can handle many tasks, people might start thinking, `Since the robot can do everything, I don’t need to help.'''}. P4 also expressed worry that this interaction could reinforce potential stereotypes: \textit{``People might start thinking that blind people can’t do anything on their own.''}
To support the effective use of robotic aids, P6 drew a parallel to guide dog training, emphasizing the need to learn when to rely on the robot: \textit{``Just like guide dog users train with their dogs, I think there’s a need to develop a skill set for interacting with the robot...Deciding when to speak for yourself and when to rely on the robot is also part of walking ability.''}
More so, as users become more familiar with the guide robot, their approach to using the technology could evolve. The quote from P8 illustrated this potential: \textit{``As you become more skilled, the way you use the robot could change. The reliance on Omakase Mode will likely decrease, and a combination of Monitor and Boss will become more common. Initially, you might use the robot’s automatic features more often due to uncertainty, but as you become more accustomed, you’ll rely less on these features, even in scenarios like waiting in line.''} 
This suggests that over time, users might develop a more balanced approach to interacting with the robot, leveraging active user control to enhance human engagement.

\section{Discussion}
The goal of this paper is to inform the field of HRI about key considerations for designing navigation robots for blind people, focusing on control frameworks for independence~\cite{lee2021designing}. Our work specifically contributes to interdisciplinary discussions on user control in autonomous navigation, which have involved efforts from human-computer interaction and robotics researchers (\eg,~\cite{zhang2023follower, ranganeni2023exploring}). Based on our findings, we discuss design implications for shared control and suggest directions for future work, while acknowledging the limitations of our study.

\subsection{Design Implications for Shared Control}

\subsubsection{Building Support for Interaction Modes of Autonomy}
Our study of human-robot interactions, leveraging different modes of autonomy, revealed key insights into shared control. One aspect was blind users receiving \textit{robot support} and negotiating through \textit{mutual support}. In navigation task sessions, blind participants actively engaged with the robot and the crowd, transitioning between Monitor Mode to assess their surroundings and Boss Mode to directly influence interactions with the crowd. Our post-session interviews further highlighted the need for this combination. Participants discussed \textit{robot support} as a means to exercise their control, using the information it provided to better navigate and interact with the crowd. This dynamic reinforced the idea that the robot facilitates, rather than overrides, human decision-making and agency (see Sec.~\ref{sec:result_user_robot_involvement}). This perspective surfaces interdependence as a critical framework to describe relationships between blind users and robots~\cite{bennett2018interdependence}. More so, participants extended this framework to \textit{mutual support}, highlighting instances in which they could assist the robot in navigating complex situations (see Sec.~\ref{sec:comment_monitor_interaction_mutual_support}). They asked for situational updates from the robot to gain the ability to explore options for addressing social navigation challenges. We could see this dynamic aligning with previous discussions on sighted guides~\cite{kamikubo2020support} and guide dogs~\cite{hwang2024towards} to realize collaborative efforts with blind users. Indeed, participants in our study referred to their interactions with sighted guides to match their interactions with the robot.

To realize such \textit{robot support} and \textit{mutual support} in autonomous navigation for blind users, incorporating scene descriptions and status announcements is a crucial step. Our observations of user queries during navigation tasks (see Sec.~\ref{sec:result_Monitor_Conversation}), such as questions about obstacles in the robot’s path, crowd density, nearby pedestrian activities, queue starting points, and user position within the line, could guide the design of detailed scene descriptions. Participant feedback (see Sec.~\ref{sec:comment_boss_interaction_robot_support}) further revealed a need for more specific information about surrounding individuals, including their distance, direction, and characteristics \eg, whether they are adults or children. To address these needs, robots should be equipped with advanced perception modules capable of detecting and identifying obstacles, recognizing pedestrians and their attributes (\eg, position, activity, quantity), and detecting lines. Additionally, supporting effective communication and negotiation is critical, with robots responding to user requests and providing accurate feedback. Leveraging Multimodel Large Language Models~\cite{GPT4o} may assist in achieving this functionality~\cite{Kaniwa2024ChitChatGuide}. 

Robots should also inform users of potential failures, such as crowd navigation issues or delays in route planning. While a simple system that announces reasons for the robot stopping has been implemented (\eg, ``There is a person ahead'')~\cite{cabot-github}, this functionality needs improvement to capture the often complex reasons behind robot stops.
This involves improving the robot's ability to interpret complex environments and clearly communicate stop or delay reasons in a way blind users can understand to ease their concerns~\cite{Hong2024Understanding}.
Addressing these challenges will require further research in both robotic perception and human factors.
A possible solution could use velocity values obtained from the robot's local planner~\cite{macenski2024smac}, the local cost map around the robot~\cite{macenski2023survey}, and images of the surroundings as inputs to a Vision-and-Language Model~\cite{GPT4o} to generate natural language explanations.

\subsubsection{Balancing User and Robot Involvement}
User involvement was found to be another critical aspect of shared control. Participants expressed a strong preference for taking an active role in their navigation, rather than following the robot entirely, as would be the case in Omakase Mode. This preference was shaped by various factors, including a desire to maintain their sense of agency and address complex navigation challenges. Social acceptance was also a key factor, particularly in scenarios where participants felt it was more appropriate to interact directly with the crowd rather than relying on the robot (see Sec.~\ref{sec:result_user_robot_involvement}). This finding connects to the literature on social accessibility (\eg,~\cite{shinohara2016self}), which notes that blind people often negotiate boundaries to avoid unwanted attention from those around them~\cite{worth2013visual}.

Yet, striking a balance between user and robot involvement was deemed critical. During navigation tasks, participants let the robots take the lead role in certain contexts, such as navigating dynamic crowds or following others in line. Interestingly, these scenarios reflect two primary challenges identified in our preliminary study. We observed that Omakase Mode, with its autonomous navigation, addressed real-world difficulties participants commonly faced.

Moving forward, it is important to consider this delicate balance of involvement in autonomous robot navigation, particularly in longitudinal contexts. Over time, users' relationships with the robots and their preferences for control may evolve. This aligns with broader discussions on user complacency with automation~\cite{merritt2019automation}. Interestingly, our findings suggest a unique trend: as participants grew accustomed to interacting with the robot, many envisioned scenarios where they would seek more control, leveraging Boss and Monitor Mode. At the same time, participants reflected on societal implications, expressing concerns about over-reliance on robots (see Sec.~\ref{sec:comment_monitor_interaction_mutual_support}). 

To address these complexities, personalization in shared control is a much-needed direction. For example, users could choose who leads the interaction---whether the robot or themselves---based on context or preferences. While external communication with surrounding pedestrians, such as non-visual alerts like beeping sounds, have been explored for blind users~\cite{kayukawa2019bbeep,rodriguez2012assisting} or other  groups~\cite{zhang2022understanding}, the use of adaptive strategies remains underexplored. 
Kayukawa \etal\cite{kayukawa2019bbeep} report that blind users would not use sound-emitting systems in environments where silence is expected, such as hospitals and libraries.
Our study offers insights into various commands, including alert or inquiry approaches, that take situational factors into account. For instance, participants favored robots taking charge in noisy, unfamiliar, or urgent situations, reinforcing the need for adaptive designs in shared control systems.

\subsection{Limitations and Future Work}

\indent \textit{Expanding Real-World Scenarios.} In this paper, based on the findings from the preliminary study, we prepared two challenging social navigation scenarios with actors and visitors in a museum during business hours. These scenarios were designed to explore the desirable mode of autonomy which varies by situations~\cite{ranganeni2023exploring}. During the user tasks, we observed four main types of crowd situations, along with more complex scenarios involving combinations of these types. For instance, there were mixes of stationary and dynamic elements, such as pedestrians walking through groups of standing individuals, or combinations of artificial and natural scenarios, like regular visitors moving aside upon hearing alerts directed at actors.
We expect more complex situations to occur in the real world, such as continuous pedestrian streams or coexistence of multiple lines. In these contexts, users' strategies may shift.
For future work, we plan to extend our studies to more challenging environments outside the museum, such as train stations and busy shopping districts, which were identified as particularly difficult for blind people in our preliminary interviews.

\textit{Revisiting Monitor Mode.} In this study, the Wizard-of-Oz approach was used during Monitor Mode, with a human experimenter simulating the robot's responses based on a predefined set of information the robot could detect, such as surrounding pedestrians and obstacles~\cite{cabot-github}. This method was adopted to encourage participants to reflect on interaction aspects of this level of autonomy, rather than focusing on technical aspects. While recent advances have brought speech synthesis to levels comparable to the human voice~\cite{shen2024naturalspeech}, limitations in reliability and user trust remain due to potential inaccuracies in responses~\cite{Kaniwa2024ChitChatGuide,clark2019makes}. 
Interestingly, in our study, participants were still inclined to issue user inquiries and alerts to nearby bystanders themselves, even though they knew the robot's voice was provided by the human experimenter. This suggests that interaction behaviors in Monitor Mode may not solely depend on the quality or content of the response but on the perceived source of assistance.
Future research is needed to understand the interaction strategies and control preferences when assistance is provided by a machine~\cite{Kaniwa2024ChitChatGuide}. The perception of assistance as ``human-driven'' versus ``machine-driven'' may influence user trust and engagement.

\textit{Exploring Long-Term Interactions and User Contexts.} Although participants were given a training session to familiarize themselves with the robot’s basic interaction (as detailed in Sec.~\ref{sec:Procedure}), some unexpectedly switched to Monitor Mode or appeared confused by the robot's temporary stops in Omakase Mode, even when there were no technical issues or unexpected external factors affecting their movement. This suggests that blind users' interactions with the robot can change based on their familiarity with its functionality and their experience level. In our study, 3 participants were completely new to robot interactions, and 2 had more exposure than others (having experienced 5-6 trials previously with the robot). For those who were using the robot for the first time, the short-term nature of these experiments only reflected initial impressions and immediate responses to first interactions. While some insights into user behavior in the longitudianl context were discussed in Sec.~\ref{sec:comment_monitor_interaction_mutual_support}, further long-term studies involving continued use are necessary to explore how preferences and interaction strategies for user control evolve over time.

On a similar note, with only two participants (P4 and P11) as guide dog users, a more representative user group might reveal different interaction preferences. Future studies should focus on participants with diverse attributes, including those accustomed to walking with robots or guide dog users, to better explore unique interactions, strategies, and challenges.

\section{Conclusion}
In this study, we explored how user autonomy and interface design in future navigation robots for blind individuals should evolve beyond \textit{omakase}, the current fully robot-led, approaches. 
Through interviews with 14 blind participants, we identified two key social navigation challenges: navigating through streams of people and following lines, along with the strategies used to address them.
We then conducted a user study where 13 blind participants navigated a museum with an autonomous robot, exploring their preferences across different modes of autonomy (Omakase, Monitor, and Boss). 
The results revealed a preference for the active Boss mode, where participants took charge of crowd interactions, rather than relying entirely on the robot. 
Participants also appreciated the supportive role of the robot, using Monitor mode to understand the environment, negotiate movement, and facilitate user-robot interactions.
These findings highlight the importance of shared control and user involvement for blind people. 
Studies on shared control for blind navigation are still neither active nor sufficient, requiring further research in these intersectional domains. 
Such studies will significantly expand the applicability of navigation robots in the future.

\begin{acks}
We thank the participants in our study as well
as the anonymous reviewers for their comments that further strengthened this paper. We also like to thank the Miraikan staff, including Hiromi Kurokawa and Takashi Suzuki, for their help in running the studies. Rie Kamikubo and Hernisa Kacorri were supported from the National Institute on Disability, Independent Living, and Rehabilitation Research (NIDILRR), ACL, HHS (grant \#90REGE0024).

\end{acks}
\balance
\bibliographystyle{ACM-Reference-Format}
\bibliography{ref,ref_findings}
\end{document}